# Anisotropic magnetic properties and giant rotating magnetocaloric effect in double-perovskite $Tb_2CoMnO_6$

J. Y. Moon,[1] M. K. Kim,[2,3] D. G. Oh,[1] J. H. Kim,[1] H. J. Shin,[1] Y. J. Choi,[1,*] and N. Lee[1,*]

[1]*Department of Physics, Yonsei University, Seoul 120-749, Korea*
[2]*Center for Correlated Electron Systems, Institute for Basic Science, Seoul 151-742, Republic of Korea*
[3]*Department of Physics and Astronomy, Seoul National University, Seoul 151-747, Republic of Korea*

We investigated the anisotropy of the magnetic and magnetocaloric properties of single-crystalline double perovskite $Tb_2CoMnO_6$, which crystallizes in a monoclinic $P2_1/n$ structure. Due to dissimilar magnetic anisotropy, the ferromagnetic order of the $Co^{2+}$ and $Mn^{4+}$ moments emerges along the *c*-axis at $T_C = 100$ K, and the larger $Tb^{3+}$ moments align perpendicular to the *c*-axis, below $T_{Tb} = 15$ K. The intricate temperature development of the metamagnetism along the *c*-axis results in a large negative change in the magnetic entropy at low temperature. On the other hand, the larger but almost reversible magnetization, perpendicular to the *c*-axis, results in a small and positive entropy change. This highly anisotropic magnetocaloric effect (MCE) leads to a giant rotational MCE, estimated to be 20.8 J/kg·K. Our findings, based on the magnetic anisotropy in $Tb_2CoMnO_6$, enrich fundamental and applied research on magnetic materials, considering the distinct magnetic characteristics of double perovskites.

[*]Electronic addresses: phylove@yonsei.ac.kr and eland@yonsei.ac.kr

# I. Introduction

One of the ideas behind the exploration of new magnetic material is the realization of the desired functional properties for potential use in a wide range of practical applications, such as high-density data storage, medical devices, and sensor technologies [1-6]. The magnetocaloric effect (MCE), described as the change in the temperature ($T$) of a material by the magnetic- field ($H$) variation, is a functional property that can realize energy-efficient magnetic refrigeration for clean technology. Recent efforts for enhancing the feasibility of magnetic refrigeration involve the development of a rotating MCE, in which the effect can be attained by rotating the refrigerant at constant $H$. Refrigerant cooling can be realized by strong magnetic anisotropy with advantages, including technical simplicity and device compactness. Rotating MCE near room-temperature, which offers potential refrigeration techniques for domestic usage and microelectronic devices, has recently been demonstrated in several alloy systems, such as $NdCo_5$ [7] (the adiabatic $T$ change, $\Delta T_{ad}$ = 1.6 K for 1.3 T at 280 K), $Er_2Fe_{14}B$ [8] ($\Delta T_{ad}$ = –0.9 K for 1.9 T at 280 K), and $NdCo_4Al$ [9] (the magnetic entropy change, $\Delta S_M$ = 1.3 J/kg·K for 1 T at 295 K). In addition, cryogenic rotating MCE, essential as a substitute for $^3He/^4He$ dilution refrigeration and hydrogen-gas liquefaction utilized as an alternative fuel, has been observed in several insulating transition metal oxides, such as $TbMnO_3$ [10] (the magnetic entropy change obtained by rotation, $\Delta S_R$ = 9.0 J/kg·K for 5 T at 15 K), $HoMn_2O_5$ [11] ($\Delta S_R$ = 12.4 J/kg·K for 7 T at 10 K), $TmFeO_3$ [12] ($\Delta S_R$ = 9.0 J/kg·K for 5 T at 17 K), $KTm(MoO_4)_2$ [13] ($\Delta S_R$ = 9.8 J/kg·K for 5 T at 10 K), and $KEr(MoO_4)_2$ [14] ($\Delta S_R$ = 13 J/kg·K for 5 T at 10 K). The exploitation of insulating oxides provides advantages, including easy manufacturability and strong stability, and prevents inefficiency due to eddy current. As the rotating MCE is varied finely by the anisotropic magnetic properties, a scientific understanding of the diverse magnetic phases and interactions is crucial for designing and discovering new compounds that exhibit the desired rotating MCE.

Double perovskite compounds with appropriate combinations of several magnetic ions have been recently investigated because of their intriguing physical properties, such as metamagnetism [15-17], exchange bias [18,19], re-entrant spin-glass state [20-23], and multiferroicity [24-28], arising from the intricate magnetic interactions and ionic valence/antisite disorders between the mixed-valence magnetic ions. In double perovskite $R_2CoMnO_6$ (R = La, …, Lu) compounds, $Co^{2+}$ and $Mn^{4+}$ ions are alternately located in corner-shared octahedral environments. The dominant $Co^{2+}$ and $Mn^{4+}$ superexchange interactions leads to a ferromagnetic order, whose transition, $T$, varies linearly from 204 K for $La_2CoMnO_6$ to 48 K for $Lu_2CoMnO_6$, as the size of the rare-earth ions decrease [29-31]. With the incorporation of a magnetic rare-earth ion, the additional ordering of the rare-earth ion at a lower $T$ results in a substantial modification of the magnetic properties. In $Er_2CoMnO_6$, for instance, the ferrimagnetic order is activated by the long-range order of the $Er^{3+}$ moments below $T_{Er}$ = 10 K, in addition to the ferromagnetic order of the $Co^{2+}$ and $Mn^{4+}$ moments arising at $T_C$ = 67 K [32,33]. Strong magnetic anisotropy and inversion of the magnetic hysteresis loop were observed with respect to the delicate balance between the magnetic rare-earth and $Co^{2+}/Mn^{4+}$ moments, under magnetic fields [34].

In this work, in order to investigate the role of magnetic rare-earth ions, and the influence of the anisotropic characteristics on the MCE in one of the double-perovskite oxides, we synthesize single crystals of double perovskite $Tb_2CoMnO_6$ (TCMO), using the conventional flux method. We explore the influence of magnetic anisotropy and metamagnetism on the rotating MCE in TCMO. The distinctive magnetic anisotropy originates from the $Co^{2+}/Mn^{4+}$ moments aligned ferromagnetically along the $c$-axis; the large $Tb^{3+}$ moments are oriented perpendicular to the $c$-axis. A large negative entropy change was observed along the $c$-axis, attributed to the intricate temperature evolution of the metamagnetic transition. An almost reversible hysteretic behavior of the isothermal magnetization perpendicular to the $c$-axis results in a small positive entropy change. As a result, the giant rotational MCE is determined to be $\Delta S_\theta$ = 20.8 J/kg·K at 2 K. Our studies on the strong anisotropic magnetic properties of double-perovskite compounds offer crucial clues for exploring suitable materials for magnetic functional applications.

## II. Experimental

Single crystals of TCMO were synthesized using the conventional flux method with $Bi_2O_3$ flux. To prepare polycrystalline TCMO samples, a stoichiometric ratio of the $Tb_4O_7$, $Co_3O_4$, and $MnO_2$ powders were mixed in a mortar. These mixed powders were pelletized and calcined at 1100 °C for 12 h. The calcined pellets were reground, pelletized, and sintered at 1200 °C for 24 h. The same grinding and sintering procedures were performed at 1300 °C for 48 h. The presintered polycrystalline powder was mixed with $Bi_2O_3$ flux at a ratio of 1:12, followed by heating to 1280 °C in a Pt crucible. It was then dissolved in the flux at the same $T$ for 5 h. The crucible was cooled slowly to 985 °C at a rate of 2 °C/h, and cooled down to room $T$ at a rate of 250 °C/h. The crystallographic structure of the TCMO crystals was confirmed using an X-ray diffractometer (D/Max 2500, Rigaku Corp., Japan). The $T$ and $H$ dependences of the DC magnetization ($M$) were obtained using a vibrating sample magnetometer at T = 2-300 K and H = –7-7 T, using a physical properties measurement system (PPMS, Quantum Design, Inc., USA). The $T$ dependence of the heat capacity ($C$) was measured using the standard relaxation method in the PPMS.

**III. Results and discussion**

To refine the crystallographic structure of TCMO, powder X-ray diffraction measurement was performed at room temperature. Figure 1(a) shows the X-ray diffraction pattern of the ground TCMO single crystals, and the simulated pattern from the Rietveld refinement using the Fullprof software; the observed and calculated patterns are depicted as open circles and solid lines, respectively. The result suggests that TCMO crystallizes in a monoclinic $P2_1/n$ double-perovskite structure with the unit cell $a$ = 5.2975 Å, $b$ = 5.6053 Å, and $c$ = 7.5470 Å, and $\beta$ = 89.9343° with good agreement factors; $\chi^2$ = 9.15, $R_P$ = 2.17%, $R_{wp}$ = 3.54%, and $R_{exp}$ = 1.17%, as summarized in Table I. As per the specific atomic positions in the unit cell listed in Table I, the $Tb^{3+}$ and $O^{2-}$ ions are located in the 4e sites, while the $Co^{2+}$ and $Mn^{4+}$ ions are located in the 2b and 2c sites, respectively. The crystal structures viewed from the $a$- and $c$-axes are depicted in Figs. 1(b) and (c), respectively. The double-perovskite structure is characterized by a rock-salt like arrangement of corner-shared octahedral units of $Co^{2+}$ and $Mn^{4+}$ ions. The oxygen octahedral cages are strongly distorted due to the small radius of the $Tb^{3+}$ ion. The distortion leads to $O^{2-}$ ion shifts, in the bonds connecting the $Co^{2+}$ and $Mn^{4+}$ ions, away from the crystallographic axes. The bond angles between the $Co^{2+}$ and $Mn^{4+}$ ions via the $O^{2-}$ ions are estimated to be 144.0249 °, 126.2522 °, and 154.4553 °, respectively, for different positions of the $O^{2-}$ ions (Table I).

To investigate the magnetic properties of TCMO, the $T$ dependence of the magnetic susceptibility, $\chi = M/H$, was measured at $\mu_0H$ = 0.2 T, on warming after zero-field-cooling (ZFC), and cooling (FC) in the same field. The anisotropic $\chi$ was obtained for the field along the $c$-axis ($H//c$), and perpendicular to the $c$-axis ($H\perp c$), respectively, as shown in Figs. 2(a) and (b). The overall $T$-dependence of $\chi$ for the two different orientations exhibits strong magnetic anisotropy, indicating that the spins are mainly aligned along the $c$-axis. For the ZFC $\chi$ in $H//c$, a minute negative magnitude of $\chi$ at 2 K was observed due to a typical remnant $H$, which remained negative on cooling. The ZFC $\chi$ with an almost fully demagnetized state at low temperature increases above approximately 20 K, due to thermally activated domain wall motions. A peak around 90 K represents an additional domain wall depinning process. The $T$ at which the ZFC and FC $\chi$ curves begin to split was observed, indicative of the onset of magnetic irreversibility. The ferromagnetic order corresponding to the dominant $Co^{2+}$ and $Mn^{4+}$ superexchange interactions emerges at $T_C$ = 100 K, which can be determined by the sharp anomaly in the $T$ derivative of $\chi$. The thermally hysteretic behavior between the ZFC and FC $\chi$'s around $T_C$ indicates the first-order nature of the transition.

The $T$ dependence of the heat capacity divided by the $T$ ($C/T$) measured at zero $H$ also shows a distinct anomaly at $T_C$ (Fig. 2(c)). As studied by the previous neutron diffraction experiments on the polycrystalline TCMO [15,35], it was suggested that exchange couplings between the $Tb^{3+}$ and $Co^{2+}/Mn^{4+}$ sublattices would be activated in the low $T$ regime below $T_C$. The short-range antiferromagnetic order of the $Tb^{3+}$ moments at low $T$ and zero $H$ was also observed as a broad peak at a low scattering angle [15,36], as revealed by an abrupt increase of $C/T$ below $T_{Tb}$ = 15 K. To evaluate the entropy change by the spin order of the $Tb^{3+}$ ions ($\Delta S_{Tb}$) under zero $H$ below $T_{Tb}$, the influence of the ferromagnetic $Co^{2+}/Mn^{4+}$ order and the interaction between the $Tb^{3+}$ and $Co^{2+}/Mn^{4+}$ sublattices on $C/T$ in the low-$T$ regime was excluded by subtracting the contribution, using the following equation:

$$C/T \sim \gamma + \rho T^{1/2} + \beta T^2, \qquad (1)$$

where $\gamma$, $\rho$, and $\beta$ are coefficients associated with the electrons, magnons, and phonons of the $Co^{2+}/Mn^{4+}$ moments, respectively. The dark-gray dashed curve indicates the fitting curves of $C/T$ due to the contribution of the ferromagnetic $Co^{2+}/Mn^{4+}$ order and the interaction between the $Tb^{3+}$ and $Co^{2+}/Mn^{4+}$ sublattices at low $T$. The estimated $\Delta S_{Tb}$ was found to be 6.8 J/mole·K, which is only 16% of the expected value of the fully saturated $Tb^{3+}$ moments, i.e., $2R \ln(2J + 1)$ = 42.7 J/mole·K, where R is the gas constant and $J$ is the total angular momentum ($J$ = 6 for the $Tb^{3+}$ ion).

The detailed magnetic anisotropy in TCMO was examined through the isothermal $M$ for the two different orientations, measured up to $\mu_0 H$ = 7 T at 2 K, as displayed in Figs. 3(a) and (b). The initial $M$ curve at $H//c$ shows a gradual increase as $H$ increases, before a sudden increase at 3.8 T. On increasing $H$ further, $M$ increases linearly with a slight slope, resulting from the reorientation of a small portion of antiferromagnetic spins due to the antisites and/or antiphase boundaries. $M$ at a maximum $H$ of 7 T is 5.81 $\mu_B$/f.u., which is nearly compatible with the summation of the $Co^{2+}$ (S = 3/2) and $Mn^{4+}$ (S = 3/2) magnetic moments in a formula unit. The consecutive sweeping of $H$ between +7 and –7 T demonstrates two sharpened transitions occurring at –1.47 and –3.46 T with an intermediate plateau. The hysteretic behavior of the full curve was observed with the remnant $M$, with $M_r$ = 4.71 $\mu_B$/f.u., and the coercive field, with $H_c$

= 1.50 T. On the other hand, $M$ at $H\perp c$ exhibits a continuous increase up to 7 T, without magnetic hysteresis. Regardless of the hard-magnetic $c$-axis for the ferromagnetic $Co^{2+}$ and $Mn^{4+}$ sublattice, the large magnetic moment of 9.73 $\mu_B$/f.u. at 7 T suggests a peculiar magnetic anisotropy for the $Tb^{3+}$ moments perpendicular to the $c$-axis. This is compatible with the disappearance of the broad scattering peak with the enhancement of the other magnetic scattering reflections under an applied $H$, indicating the development of a long-range order of $Tb^{3+}$ moments in the $ab$ plane, in previous neutron scattering studies [15,36].

The $T$ evolution of the anisotropic characteristics of $M$ up to 7 T ($T$ = 5, 10, 40, and 100 K) is depicted in Fig. 4. At 5 K and $H//c$, the double-step metamagnetic transitions broaden, while the coercive field appears to be slightly enhanced as $H_c$ = 1.82 T (Fig. 4(a)). At 10 K, the area within the magnetic hysteresis loop along $H//c$ is considerably reduced with the decrease in $H_c$ ( $H_c$ = 1.10 T), but the remnant $M$ is almost maintained ($M_r$ = 4.45 $\mu_B$/f.u.) (Fig. 4(b)). At 40 K, the metamagnetic transitions continue to remain but the magnetic hysteresis becomes narrow with $M_r$ = 1.86 $\mu_B$/f.u and $H_c$ = 0.24 T (Fig. 4(c)). At 100 K, a slight hysteretic behavior remains but the metamagnetic transitions disappear (Fig. 4(d)). At $H\perp c$, a very narrow but large magnetic hysteresis loop is observed at 5 K, and the loop is enlarged at 10 K (Figs. 4(e) and (f)). As $T$ increases further, the area of the magnetic hysteresis loop rapidly shrinks, and $M_r$ and $H_c$ also decrease (Figs. 4(g) and (h)).

Due to the contrasting magnetic properties for the two different orientations in the TCMO, a strongly anisotropic MCE was attained by measuring the initial $M$ curves with dense $T$ steps ranging from 2-150 K, in Fig. 5. At $H//c$, the sharp feature of the metamagnetic transition at 2 K moves gradually to a lower $H$ and becomes broader as $T$ increases (Fig. 5(a)). In contrast to the typical reduction of the $M$ values with the increase in $T$, the $M$ value in a given $H$ regime is larger than that at lower $T$. As $T$ is further increased, the degree of shift toward a lower $H$ is reduced, and the $M$ value at 7 T gradually decreases. As shown in Fig. 5(b), this tendency of the initial $M$ curves alters at approximately 70 K; thus, the $M$ value exhibits a typical decreasing trend in most of the $H$ regimes, as $T$ increases. At $H\perp c$, the slope of the isothermal $M$ curve at the initial $H$ regime is the greatest at 2 K, but it decreases rapidly around 2 T, followed by the monotonous increment of $M$ up to 7 T. As $T$ is increased to $T_{Tb}$, the initial slope of $M$ progressively reduces, while the maximum $M$ value at 7 T continuously increases, leading to the intersection of the $M$ curves at approximately 5 T (Fig. 5(c)). Above $T_{Tb}$, the overall magnitude of $M$ is reduced in most of the regimes of $H$ with the increase in $T$, as shown in Fig. 5(d).

The MCE in TCMO can be quantified by estimating the isothermal magnetic entropy change ($\Delta S_M$) at a given $T$, derived from Maxwell's relation:

$$\Delta S_M(T,H) = -\mu_0 \int_0^{H_f} \frac{\partial M(T,H)}{\partial T} dH, \qquad (2)$$

where $\mu_0$ is the magnetic permeability in vacuum and $H_f$ is the end point of $H$ for the integral ($H_f$ = 3, 5, and 7 T); $T$, which the gradient of $M$, $\frac{\partial M(T,H)}{\partial T}$, was calculated approximately from two adjacent data points. The $T$ dependence of the calculated $\Delta S_M(T)$ for $H//c$ and $H \perp c$ is plotted in Figs. 6 (a) and (b), respectively, for $H$ regimes of $\Delta H$ = 0-3, 0-5, and 0-7 T, respectively. The $\Delta S_M$ values for both orientations exhibit broad peaks at $T_C$, where the maximum $\Delta S_M$ values for $\Delta H$ = 0-7 T were found to be 5.8 and 2.9 J/kg·K, respectively, for $H//c$ and $H \perp c$. The considerably larger magnitude of $\Delta S_M$ at $T_C$ for $H//c$ describes the magnetic easy $c$-axis with respect to the ferromagnetic order of the $Co^{2+}$ and $Mn^{4+}$ moments. At $H//c$, the reversed order of the $M$ magnitudes, as depicted in Fig. 5(a), result in a negative value of $\Delta S_M$. $\Delta S_M$ for $\Delta H$ = 0-7 T at 2 K is estimated to be –17.3 J/kg·K, and the magnitude of $\Delta S_M$ is reduces steeply up to $T_{Tb}$ (Fig. 6(a)). Above $T_{Tb}$, $\Delta S_M$ changes smoothly to a positive value, crossing the zero point at approximately 50 K. At $H \perp c$, the intercrossed isothermal $M$ values below $T_{Tb}$ (Fig. 5(c)) engender a substantial cancellation of $\Delta S_M$. As a result, $\Delta S_M$ for $\Delta H$ = 0-7 T is estimated to be only 1.9 J/kg·K at 2 K (Fig. 6(b)). As $T$ is further increased, $\Delta S_M$ continues to increase and reveals a broad peak around $T_{Tb}$ with a maximum $\Delta S_M$ value of approximately 6.9 J/kg·K. Above $T_{Tb}$, there is a near-absence of the estimated loss of $\Delta S_M$ but the lesser decrease of the $M$ values, upon increasing $T$, generates a continuous decrease in $\Delta S_M$.

Exploiting the distinctive characteristics of the anisotropic MCE in double perovskite TCMO compounds, the rotating MCE was measured by the angular dependence of $\Delta S_M$, denoted by $\Delta S_\theta$, where θ is the angle of deviation from the $c$-axis, i.e., $\theta = 0°$ for $H//c$ and $\theta = 90°$ for $H \perp c$ (Inset of Fig. 7(b)). Figures 7(a) and (b) display the $\Delta S_\theta$ obtained at T = 2, 5, 10, and 15 K, for 5 and 7 T, respectively. As there is a dissimilar $T$ dependence of $\Delta S_M$ between $H//c$ and $H \perp c$ (Fig. 6), the angle-dependent modulation of $\Delta S_\theta$ changes considerably with $T$. For 5 T (Fig. 7(a)), $\Delta S_\theta$ at 2 K varies considerably with the rotation of $\theta$ to 15°, and increases monotonously above 15°. A giant rotational MCE as the maximum change of $\Delta S_\theta$ = 20.8 J/kg·K was observed, which would be beneficial for rotary magnetic refrigerator technology. At 5 K, the continued variation of $\Delta S_\theta$ by the $\theta$ rotation generates a maximum $\Delta S_\theta$ of 18.3 J/kg·K. As $T$ is further increased, the gradual increase of $\Delta S_\theta$ with the rotation of $\theta$ results in a maximum $\Delta S_\theta$ of 11.8 and 7.3 J/kg·K, respectively, at 10 and 15 K. For 7 T (Fig. 7(b)), $\Delta S_\theta$ at 2 K varies linearly to 30° followed by a sudden increase to 45°, and changes negligibly above 45°. The maximum $\Delta S_\theta$ was evaluated to be 20.5 J/kg·K, comparable with the value observed at 5 T. Upon increasing $T$, the maximum $\Delta S_\theta$ progressively reduces to 18.5, 11.4, and 9.2 J/kg·K, respectively, for 5, 10, and 15 K. To check the potential of our TCMO as magnetic cryo-refrigerant using rotation, we have also calculated rotational refrigerant capacity (RC$_R$) [13,37] between the two anisotropic orientations. The RC$_R$ values in 5 and 7 T were estimated as 142 and 138 J/kg, respectively. The result is comparable with RC$_R \approx$ 170 J/kg in 5 T, estimated in the single crystalline KTm(MoO$_4$)$_2$ for the orientations between $a$ and $b$ axes [13].

## IV. Conclusion

In summary, we synthesized single crystals of double perovskite Tb$_2$CoMnO$_6$, and investigated its anisotropic magnetic properties extensively. The dominant Co$^{2+}$ and Mn$^{4+}$ superexchange interactions resulted in a ferromagnetic order below $T_C$ = 100 K, aligned mainly along the $c$-axis. The Tb$^{3+}$ moments tended to align perpendicular to the $c$-axis, below $T_{Tb}$ = 15 K. The large negative value of the magnetic entropy change along the $c$-axis was caused by the strong temperature dependence of the magnetic hysteresis and metamagnetic transition. On the contrary, the isothermal magnetization perpendicular to the $c$-axis exhibited an almost reversible hysteretic behavior, contributing to a small positive value of the magnetic entropy change. Consequently, the highly-anisotropic entropy change produced a giant rotational MCE, estimated to be 20.8 J/kg·K at 2 K. These results can motivate fundamental and applied research on magnetic materials, with focus on the magnetic characteristics of complex magnetic oxides with mixed magnetic ions.

**Acknowledgements**

This work was supported by the NRF Grant (NRF-2016R1C1B2013709, NRF-2017K2A9A2A08000278, 2017R1A5A1014862 (SRC program: vdWMRC center), and NRF-2018R1C1B6006859). J.Y.M. acknowledges the tuition support from the Hyundai Motor Chung Mong-Koo Foundation.

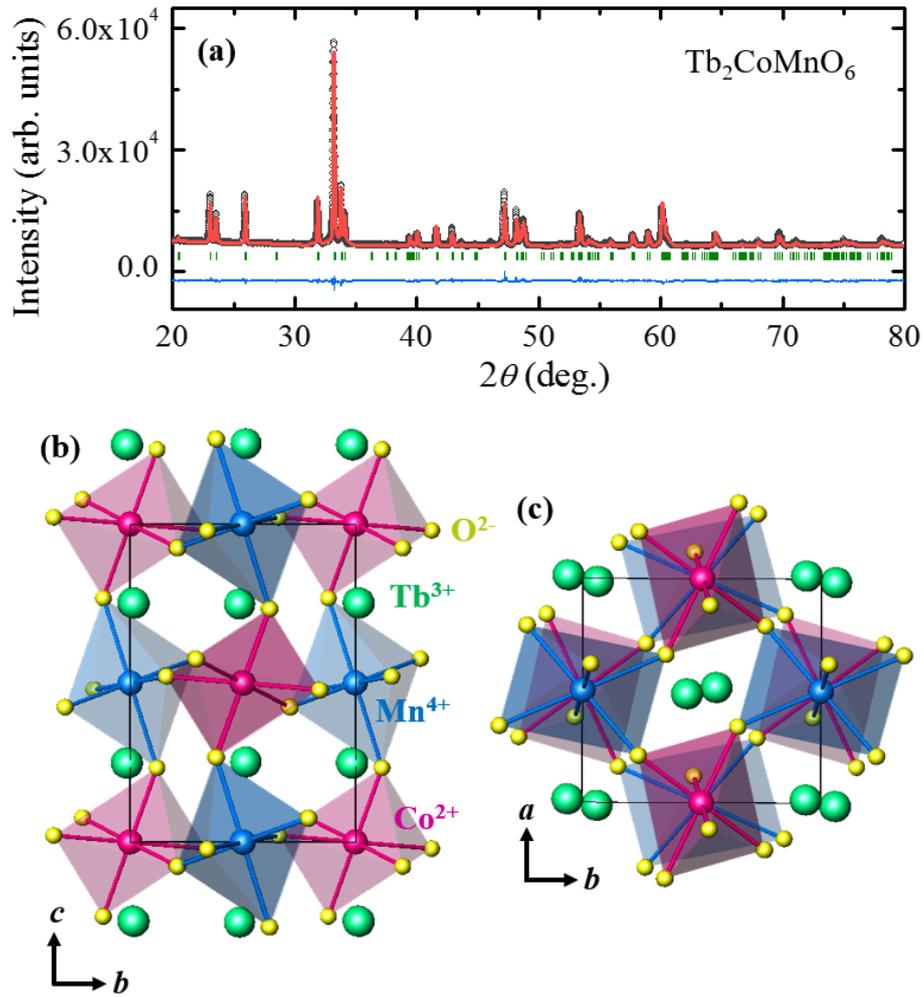

FIG. 1. (a) Observed (open circles) and calculated (solid line) powder X-ray diffraction patterns for the ground Tb$_2$CoMnO$_6$ (TCMO) single crystals. The green short lines denote the Bragg positions, and the blue curve indicates the difference between the observed and calculated patterns. (b) and (c) View of the crystallographic structure of TCMO from the (a) *a*-axis and (b) *c*-axis. The green, red, blue and yellow spheres represent the Tb$^{3+}$, Co$^{2+}$, Mn$^{4+}$, and O$^{2-}$ ions, respectively. The black box with the cross-section rectangles depicts a crystallographic unit cell.

TABLE I. Unit cell parameters, reliability factors, positional parameters, and bond lengths and angles for TCMO.

| Chemical formula | $Tb_2CoMnO_6$ |
|---|---|
| Structure | Monoclinic |
| Space group | $P2_1/n$ |
| $a$ (Å) | 5.2975 |
| $b$ (Å) | 5.6053 |
| $c$ (Å) | 7.5470 |
| $\beta$ (deg.) | 89.9343 |
| $R_p$ (%) | 2.17 |
| $R_{wp}$ (%) | 3.54 |
| $R_{exp}$ (%) | 1.17 |
| $\chi^2$ | 9.15 |
| Tb (x, y, z) | ( -0.01309, 0.06317, 0.25424 ) |
| Co (x, y, z) | ( 0.5, 0, 0 ) |
| Mn (x, y, z) | ( 0, 0.5, 0 ) |
| O1 (x, y, z) | ( 0.09342, 0.45956, 0.21537 ) |
| O2 (x, y, z) | ( 0.67052, 0.35268, 0.04280 ) |
| O3 (x, y, z) | ( 0.72099, 0.26602, 0.43497 ) |

| Bond length (Å) | | Bond angle (deg.) | |
|---|---|---|---|
| Tb – O1 | 0.6614 | | |
| Tb – O2 | 2.5731 | Co – O1– Mn | 144.0249 |
| Tb – O3 | 1.9794 | Co – O2– Mn | 126.2522 |
| Co – O1 | 2.7503 | Co – O3– Mn | 154.4553 |
| Co – O2 | 2.2912 | Tb – O1– Tb | 105.4285 |
| Co – O3 | 1.8160 | Tb – O2– Tb | 115.6135 |
| Mn – O1 | 2.0941 | Tb – O3– Tb | 142.3729 |
| Mn – O2 | 2.0295 | | |
| Mn – O3 | 2.1370 | | |

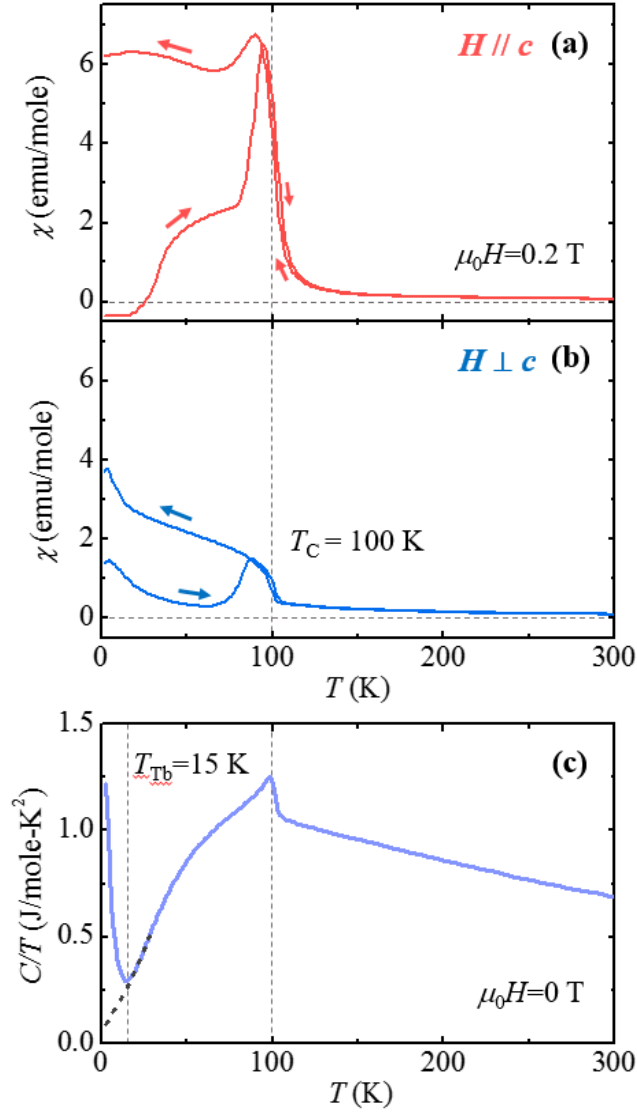

FIG. 2. (a) and (b) Temperature dependence of the magnetic susceptibility, $\chi = M/H$, of single-crystalline TCMO at $\mu_0 H = 0.2$ T, measured on heating from 2-300 K, after zero-field cooling (ZFC), and upon cooling at the same field (FC), for directions along ($H//c$) and perpendicular ($H \perp c$) to the *c*-axis, respectively. (c) Temperature dependence of the heat capacity divided by the temperature ($C/T$), measured in a zero magnetic field at $T = 2$-300 K. The dark-gray dashed curve was obtained by fitting, considering the influence of the ferromagnetic $Co^{2+}/Mn^{4+}$ order on $C/T$ at a low temperature regime. The vertical dashed lines denote the orders of the $Tb^{3+}$ ($T_{Tb} = 15$ K) and $Co^{2+}/Mn^{4+}$ ($T_C = 100$ K) moments.

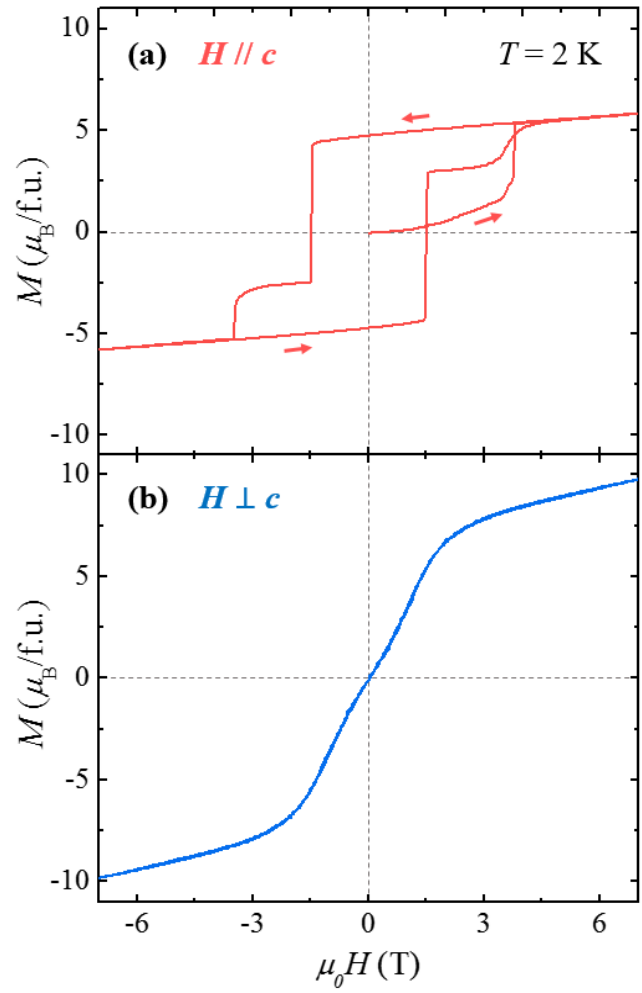

FIG. 3. Full magnetic hysteresis curves of the isothermal magnetizations measured at 2 K up to $\mu_0H = $ 7 T for (a) $H//c$ and (b) $H\perp c$.

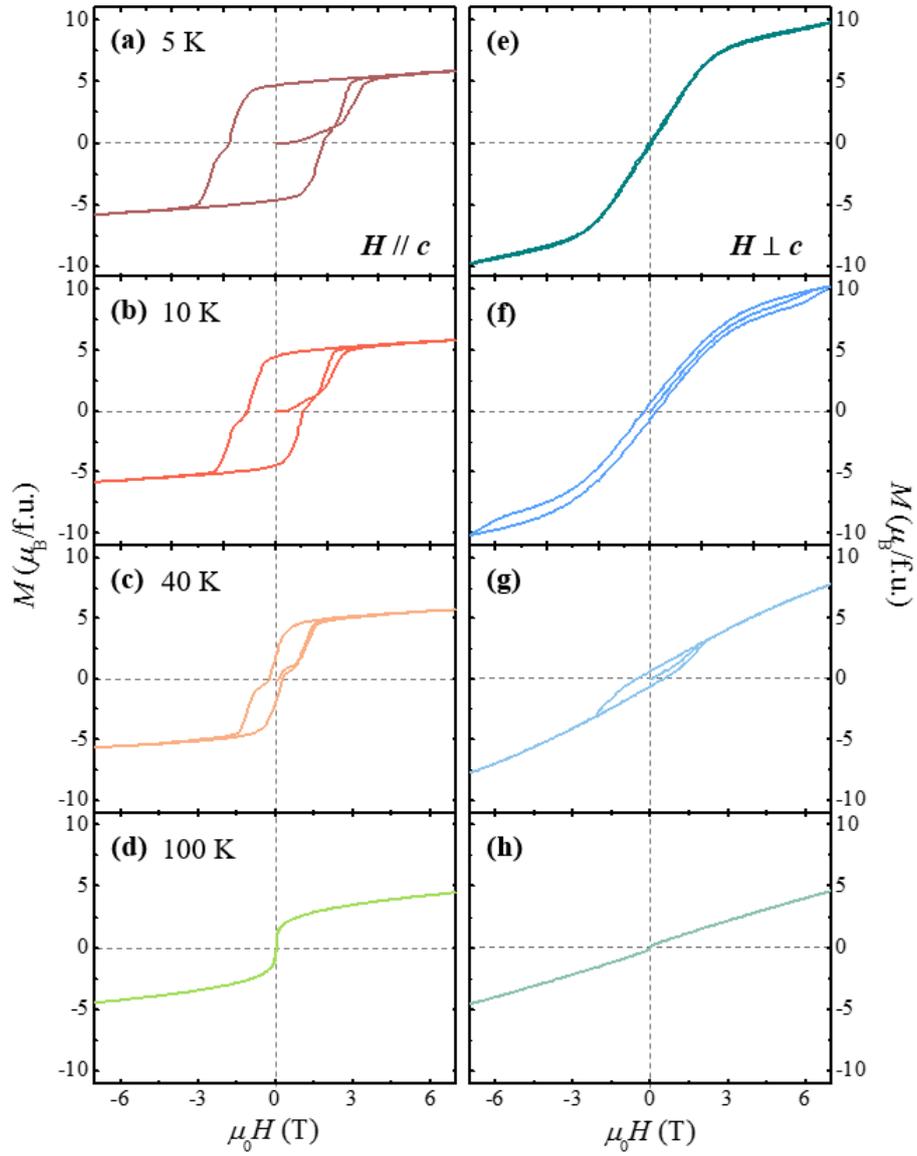

FIG. 4. Isothermal magnetizations: (a)-(d) parallel and (e)-(h) perpendicular to the *c*-axis, measured at $T$ = 5, 10, 40 and 100 K, respectively, up to $\mu_0H$ = 7 T.

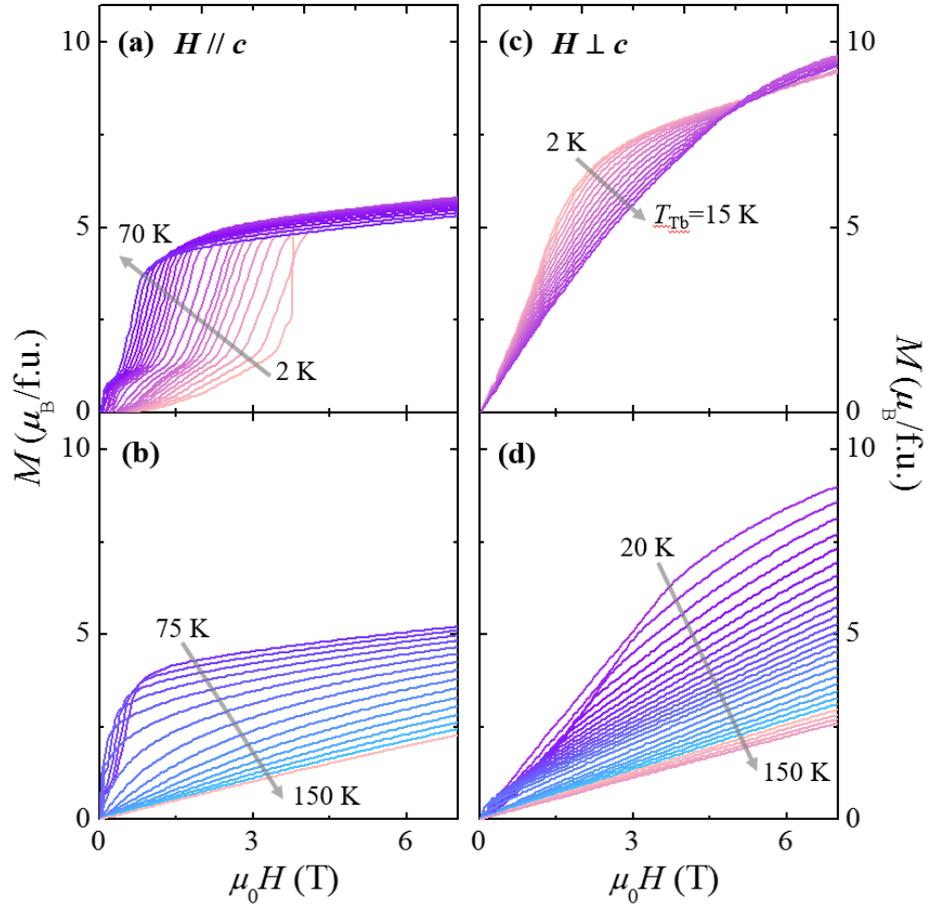

FIG. 5. Initial curves of the isothermal magnetization at $H//c$ for various temperatures: (a) $T$ = 2-70 K and (b) $T$ = 75-150 K. Initial curves of the isothermal magnetization at $H\perp c$ for various temperatures: (c) $T$ = 2-15 K and (d) $T$ = 20-150 K.

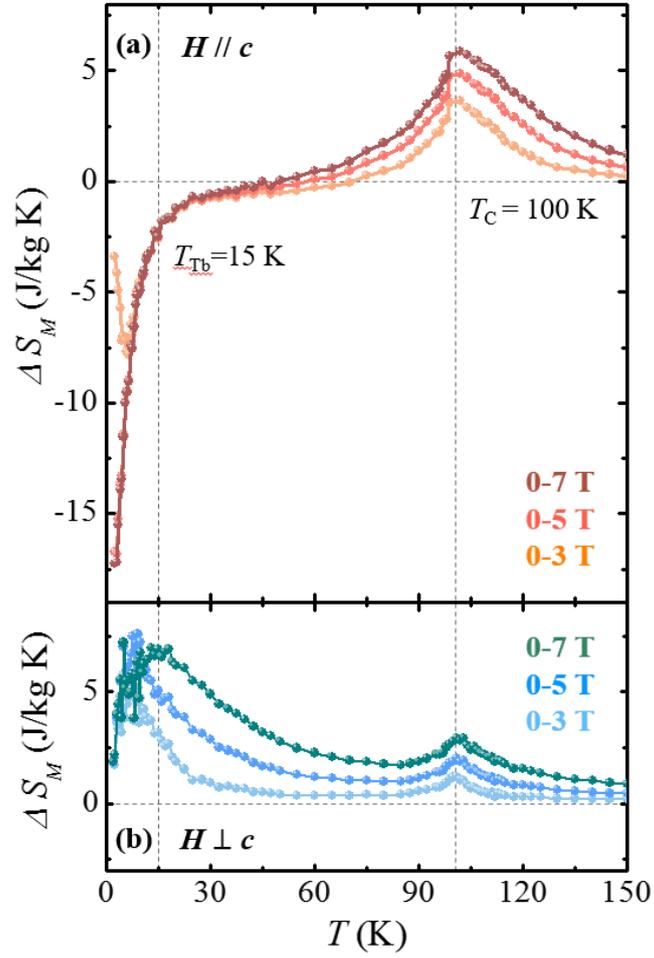

FIG. 6. Temperature dependence of the magnetic entropy change, $\Delta S_M$, obtained by integrating the temperature gradient of the initial magnetization curves in Fig. 5 for (a) $H//c$ and (b) $H\perp c$ with magnetic field regimes of $\Delta H$ = 0-3 T, 0-5 T, and 0-7 T, respectively.

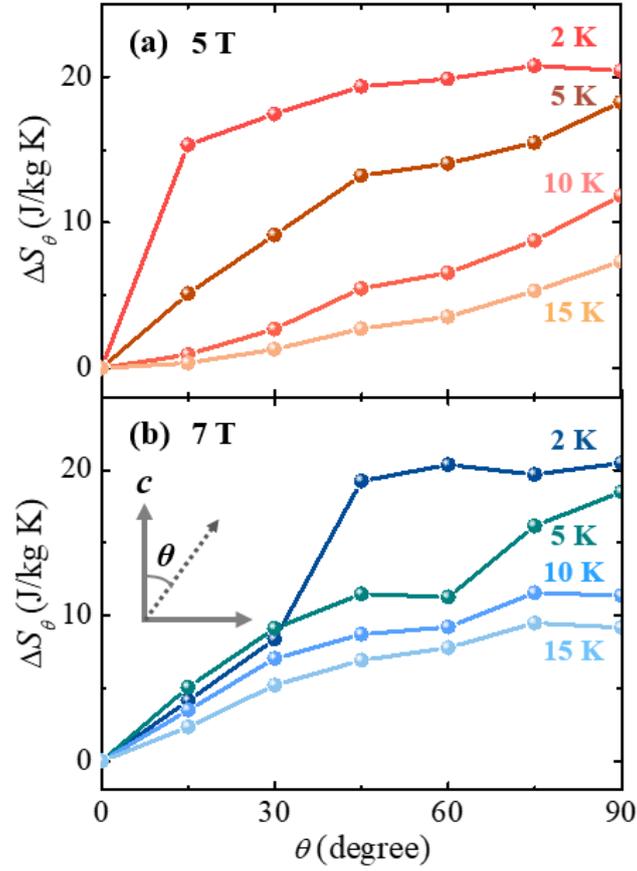

FIG. 7. Angular dependence of the magnetic entropy change, $\Delta S_\theta$, at $T = 2, 5, 10$ and $15$ K, respectively, with $\mu_0 H$ at (a) 5 T and (b) 7 T. $\theta$ is the angle deviating from the $c$-axis, i.e., $\theta = 0°$ for $H//c$ and $90°$ for $H \perp c$.